\documentclass[traditabstract]{aa} 
\usepackage{graphicx}
\usepackage{aalongtable}
\usepackage{txfonts}
\usepackage{natbib}
\usepackage{rotating}
\usepackage{lscape}
\newcommand{\gsim}{\;\lower.6ex\hbox{$\sim$}\kern-7.75pt\raise.65ex\hbox{$>$}\;}
\newcommand{\lsim}{\;\lower.6ex\hbox{$\sim$}\kern-7.75pt\raise.65ex\hbox{$<$}\;}

\newcommand{\vk}{$(V-K)$}

\begin{document}
\title{Helium in first and second-generation stars in Globular Clusters from
spectroscopy of red giants\thanks{Based on observations  collected at ESO telescopes under
programmes 072.D-507 and 073.D-0211}
}

\author{
Angela Bragaglia\inst{1}
\and
Eugenio Carretta\inst{1}
\and
Raffaele Gratton\inst{2}
\and
Valentina D'Orazi\inst{2}
\and
Santi Cassisi\inst{3}
\and
Sara Lucatello\inst{2,4,5}
}

\authorrunning{A. Bragaglia et al.}
\titlerunning{He in RGB stars in globular clusters}


\institute{
INAF-Osservatorio Astronomico di Bologna, Via Ranzani 1, I-40127
 Bologna, Italy\\
\email{angela.bragaglia@oabo.inaf.it, eugenio.carretta@oabo.inaf.it}
\and
INAF-Osservatorio Astronomico di Padova, Vicolo dell'Osservatorio 5, I-35122
 Padova, Italy\\
 \email{raffaele.gratton@oapd.inaf.it, valentina.dorazi@oapd.inaf.it, sara.lucatello@oapd.inaf.it}
\and
INAF-Osservatorio Astronomico di Collurania, via M. Maggini, I-64100 Teramo,
Italy\\
\email{cassisi@oa-teramo.inaf.it}
\and
Excellence Cluster Universe, Technische Universit\"at M\"unchen, 
 Boltzmannstr. 2, D-85748, Garching, Germany 
\and
Max-Planck-Institut f\" ur Astrophysik, D-85741 Garching, Germany
  }

\date{}

\abstract{The stars in a  Globular cluster (GC)  have always been considered
coeval and of the same metallicity. Recently, this assumption has been
challenged on the basis of  spectroscopic and photometric observations, which
show the existence of various generations of stars in GCs, differing in the
abundances of products of H-burning at high temperatures. The main final product
of this burning is He. It is then important to study the connections between
stars properties and He content.  We consider here the about 1400 stars on the
Red Giant Branch (RGB) observed with FLAMES@VLT in 19 Galactic Globular
Clusters  (GCs) in the course of our project on the Na-O anticorrelation. Stars
with different He are expected to have different temperatures (i.e., different
colours),  slightly different metallicities [Fe/H], and different luminosity
levels of the RGB bump. All these differences are small, but our study has the
necessary precision, good statistics, and homogeneity to detect them. Besides
considering the observed colours and the temperatures and metallicities
determined in our survey,  we computed suitable sets of   stellar models - fully
consistent with those present in the BaSTI   archive - for various assumptions
about the initial helium content. We find that differences in observable
quantities that can be attributed to  variations in He content are generally
detectable between stars of the Primordial (P, first-generation) and Extreme (E,
second-generation) populations, but not between the Primordial and Intermediate
ones (I).  The only exception, where  differences are significant also between P
and I populations, is the cluster NGC~2808, where three populations are clearly
separated also on the Main Sequence and possibly on the Horizontal Branch. The
average enhancement in the He mass fraction Y between P and E stars is about
0.05-0.11, depending on the assumptions.   The differences in Y, for NGC~2808
alone, are of about 0.11-0.14 between  P and I stars, and about 0.15-0.19 
between P and E stars, again depending on the assumptions.  When we consider the
RGB bump luminosity of first and second-generation  stars we find different
levels; the implied   Y difference is more difficult to quantify, but is in
agreement with the other determinations. }
\keywords{Stars: abundances -- Stars: atmospheres --
Stars: Population II -- Galaxy: globular clusters -- Galaxy: globular
clusters: individual: NGC~104 (47 Tuc), NGC~288, NGC~1904 (M~79), NGC~2808, 
NGC~3201, NGC~4590
(M~68), NGC~5904 (M~5), NGC~6121 (M~4), NGC~6171 (M~107), NGC~6218 (M~12), 
NGC~6254 (M~10), NGC~6388,
NGC~6397, NGC~6441, NGC~6752, NGC~6809 (M~55), NGC~6838 (M~71), NGC~7078 (M~15), 
NGC~7099 (M~30)} 

\maketitle

\section{Introduction}\label{intro}
Recent progresses have indicated
that large star-to-star variations exist in the He content among
stars in globular clusters (GCs), in spite of difficulties in direct derivation of He abundances  \citep[see however][]{villanova09,moehler07}, 
The first clear evidence has been obtained from
observations of multiple main sequences in \object{$\omega$~Cen}  \citep{bedin04}
and NGC~2808 \citep{dantona05,piotto07}. In $\omega$~Cen,
stars on the bluest main sequence (MS) are more metal-rich than those of the
reddest one \citep{piotto05}, a fact that can only be explained by an
He content higher by about 0.10 to 0.15 in the mass fraction
Y \citep[as originally suggested by][]{norris04} with respect to a canonical (primordial) He content of about 0.245
commonly adopted for the bulk of the stellar population in this cluster. 
The small scatter found in NGC~2808 among red giant
branch stars also implies a larger He content for the bluer MS, by roughly
the same amount. Two other clusters have been found to have wide MSs:
47 Tuc \citep{anderson09} and  NGC~6752 
\citep[][who also suggest a possible split]{milone10}.
\footnote{The split subgiant branches 
(SGB, e.g., NGC~6388: \citealt{moretti09}; \object{NGC~1851}:
\citealt{milone08}) seem more related to differences in age or CNO content. A spread in age is
probably the cause of the spread/split MS turn-off's found in many intermediate-age
Magellanic Clouds GCs (\citealt{mackey,milone09} in LMC;  \citealt{glatt}
in SMC) even if
for them -but notably not for the old Galactic GCs-  there is an alternative physical
interpretation based on the observational effects associated to the occurrence of
stellar rotation as discussed by \cite{selma}. }

A star-to-star spread in the He abundance may explain many aspects of the
horizontal branches (HB) of GCs, as extensively discussed in 
\cite{gratton10},
but we are aware that
also the case for a null spread of He in GCs is maintained in literature 
\citep[e.g.,][]{catelan}. 
Internal  variations in He are likely connected with
other chemical signatures found in GCs, related to H burning
occurring at high temperature. This has been first proposed in a modern way
by \cite{dantona02}, who noticed that the smaller masses of He-rich stars may
naturally explain the correlation between the presence of stars at the blue
extreme of the horizontal branch, and of stars extremely depleted in O in
various clusters. \cite{lee05} suggested that a super-He rich population could
explain the extreme HB stars in GCs, specifically presenting the cases of  $\omega$
Cen and NGC~2808 (where they predicted a spread or a split of the MS, just before it
was actually found). \cite{kaviraj} suggested that the far-UV properties of
GCs in the Virgo elliptical M87 could be explained by a super-He rich population
similar to the one in $\omega$ Cen.
\cite{dac08}  tried to
explain the whole distribution of stars along the HB of various clusters
(including NGC~2808, M~13, and M~3) as due to star-to-star variations in the
He content (but see also \citealt{castellani05}). 
In a companion paper \citep{gratton10}, we make a more
systematic comparison of the extremes of the distribution of stars along
the HB and of the Na-O anticorrelation. 
We find that median HB colours are quite well reproduced by a combination
of metallicity, age, and a simple mass loss proportional to metallicity.
However, the low-mass extreme of the HB is clearly correlated with the extension
of the Na-O anticorrelation \citep[see also][]{carretta09a} which, in turn,
seems to be linked with the cluster absolute magnitude, a proxy for cluster mass.

These correlations suggest that the multiple generation scenario, needed to
explain the Na-O anticorrelation,\footnote{It is generally assumed that 
first-generation stars have chemical composition similar to field stars of
similar age and metallicity (i.e., high O and Mg, low Na and
Al, etc.), while second-generation stars, formed from material polluted by
matter lost from the primordial population, show the abundance patterns
(e.g., Na-O anticorrelation) that are unique of GCs. 
} is able to explain many peculiarities of GCs. As we have seen, a crucial datum
in this scenario is the He content of second-generation stars. While the most
spectacular evidence of a variation of the He content comes from the HB and the
MS, also the red giant branch (RGB) stars are very useful because chemical
composition (but -alas- not He content) can be determined easily enough for
these stars using spectroscopy, unlike faint MS stars. This allows us to cross
correlate the abundances of various elements, providing important constraints on
the properties of the He producer, which have not been clearly identified  yet
(two main possible mechanisms have been proposed, intermediate/massive 
asymptotic giant stars -AGB- and fast rotating massive stars  -FRMS-
see e.g., \citealt{dav07} and \citealt{decressin},  respectively).
Various properties of RGB stars
are affected by the variable He abundances:
\begin{enumerate}
\item The temperature (at a given luminosity) of He-rich RGB stars is
slightly warmer than that of He-poor ones \citep[see e.g.,][]{dantona02}.
The difference is not large (only a few tens of K), and very accurate
relative determinations are required. However, both colours and line
excitation can be used to show that this really occurs.
\item The gravity (again at a given luminosity) is also different, due to
both the smaller mass and higher temperature of He-rich stars. He-rich
stars have slightly lower gravity, but this effect is very small, and
difficult to show up, because of several complications (for instance, the larger
molecular weight and reduced continuum opacity simulate a higher gravity,
offsetting most of the effect expected on e.g., ionisation equilibrium)
so we will not consider it here.
\item The [Fe/H] value is different, because 
of the different H content (broadly speaking, if Y increases, X decreases, hence
[Fe/H] increases;  
things are of course more complicated, see Sec.~\ref{deltamet}):
this difference is small but not negligible. We already noticed this effect in 
NGC~2808 \citep[][Paper I]{carretta06}.
\item Finally, the luminosity of the RGB bump is also different: the RGB bump 
is more luminous for He-rich stars \citep{iben,salaris06}, even if also in this case 
the situation is more complex (see Sect.~\ref{bump}). We showed
that possible evidence for this effect can be traced in the run of Na
abundances along the RGB of NGC~6218 (=M~12) and NGC~6752 
\citep[][Paper IV]{carretta07b}.
\end{enumerate}

In this paper, we intend to look for evidence of a variable He content in
the sample of more than 1,400 stars along the RGB of 19 GCs
for which we acquired FLAMES/GIRAFFE spectra in the last few years
\citep[presented in a series of  papers: see][-Paper VII and VIII- and references 
therein]{carretta09a,carretta09b}. 
Several features of this data set make it very promising for the
present purpose: the sample is very large; it has been analysed using a
very homogenous technique; star-to-star errors in effective temperatures,
usually the main source of errors in this analysis, were reduced to very
small values by a carefully tailored technique. In addition, we have
determined Na and O abundances for almost all these stars, allowing to
classify them into three different groups according to Na enhancement and 
O depletion (Paper VII). We called these groups P=primordial, that is stars with
composition similar to that of field stars, and likely belonging to the
primordial population; I=intermediate, and E=extreme, that are stars
with different degrees of O
depletion and Na (and likely also He) enhancement. We recall here that we
conservatively applied this definition only to stars with both O and Na
measured, thus reducing our sample to about 960 objects, the ones that will be
used in the present paper. On
the other hand, the way we analysed stars should be carefully considered
when we try to extract information from our data set. The most significant
assumptions we made are that stars of the same luminosity have the same
effective temperature and surface gravity. Whenever needed, we will
consider explicitly the consequences of these assumptions, and we will
correct our results accordingly.

Beside observations, determination of the impact of different He abundances requires
appropriate modelling. To this purpose we used a set of BaSTI evolutionary  models
\citep{pietrinferni04,pietrinferni06,pietrinferni09} purposely computed for this paper employing
two sets of heavy elements mixtures (see Sect.~\ref{sec2} and \ref{bump}  for 
details and relations).  

The structure of this paper is as follows: in Sec.~\ref{sec2} we present the observed
differences in colour, temperature, and metallicity and derive the implied
differences in He content; in Sec.~\ref{bump} we compare the RGB bumps of first and
second-generation stars, deducing a different He abundance also in this case.
Finally, we summarise and discuss our results in Sec.~\ref{ultima}.

\section{Difference in He from RGB stars}\label{sec2}
We may find different He values for the three populations using various
indicators. To actually compute the implied Y (as in the definition X+Y+Z=1 for
the chemical composition of stellar models on mass fraction), we used  relations
based on the BaSTI stellar evolutionary models. We have computed an extended set
of low-mass stellar models for various assumptions about the initial He content.
All the stellar model predictions adopted in the present analysis have been obtained
exactly in the same physical framework used for the BaSTI\footnote{The whole set of stellar models used in present work, as
well as additional predictions for He-enhanced models can be retrieved from the
URL site: {\tt http://www.oa-teramo.inaf.it/BASTI}} stellar model
library \citep{pietrinferni04,pietrinferni06}. Thus,  they are fully consistent with
the predictions corresponding to a \lq{canonical}\rq\ assumption about the
initial He abundance ($0.245\le{Y}\le0.27$).
 In particular, we used two
assumptions for the heavy element mixtures (others may be valid, but these
represent two extremes): 
\begin{itemize}
\item[$(a)$] one where the second-generation I, E stars have the same heavy
elements distribution of the first-generation P stars, both $\alpha$-enhanced
\citep[for more details see][]{pietrinferni06}  and without peculiarities in the
distribution of C, N, O, Na (a simplified approach); 
\item[$(b)$] a second one where the I, E stars have a peculiar distribution of
metals (i.e., showing a signature of the Na-O anticorrelation) and different
from the standard one of the P stars. In particular, we adopted the
"CNONa-extreme" chemical composition accounted for by   \cite{pietrinferni09},
which has the C+N+O sum enhanced by a factor of two.
\end{itemize}

It is worth mentioning that, for a fixed iron content [Fe/H], the global  
metallicity Z is higher for case $b$ than for case $a$  mainly  because the sum
of (C+N+O) is enhanced with respect the \lq {normal}\rq\ $\alpha-$enhanced
mixture. This has some implications: for the same He,  the temperature of a
CNONa-extreme track is lower than the corresponding normal $\alpha$-enhanced and
its RGB bump is fainter (see Sect. 3). This also means that we cannot give a
simple and straightforward interpretation of differences e.g., in colour or RGB
bump brightness as differences in He because the result also depends on the
mixture of heavy elements we attribute to the second-generation stars.

\begin{table*}
\centering
\caption{Difference in colour and metallicity between different populations in the 19 GCs.}
\begin{tabular}{rr rrr rrr rr rr r}
\hline\hline
NGC  &[Fe/H]
     &\multicolumn{3}{c}{$(V-K)_P$} &\multicolumn{3}{c}{$(V-K)_I -(V-K)_P$}
     &\multicolumn{2}{c}{[Fe/H]$_P$}     & \multicolumn{2}{c}{[Fe/H]$_I$} 
     &$\Delta$[Fe/H]  \\
     &Pap VIII &stars  &mean &rms &stars  &mean &rms &mean &rms     &mean &rms & \\
\hline
\object{NGC 104}  &-0.768 &24 & 0.033 &0.061 &52 &-0.010 &0.038 &-0.760 &0.041 &-0.748 &0.039 & 0.012\\  
\object{NGC 288}  &-1.305 &20 &-0.004 &0.053 &41 &-0.013 &0.048 &-1.230 &0.046 &-1.223 &0.052 & 0.007\\  
\object{NGC1904}  &-1.579 &17 &-0.010 &0.050 &16 &-0.008 &0.063 &-1.549 &0.042 &-1.536 &0.042 & 0.013\\  
\object{NGC2808}  &-1.151 &20 & 0.001 &0.059 &11 &-0.046 &0.027 &-1.167 &0.033 &-1.097 &0.056 & 0.070\\  
\object{NGC3201}  &-1.512 &31 &-0.017 &0.063 &54 & 0.003 &0.063 &-1.468 &0.061 &-1.494 &0.053 &-0.026\\  
\object{NGC4590}  &-2.265 &18 &-0.014 &0.043 &28 & 0.002 &0.040 &-2.215 &0.065 &-2.236 &0.052 &-0.021\\  
\object{NGC5904}  &-1.340 &21 & 0.002 &0.046 &53 &-0.006 &0.043 &-1.342 &0.042 &-1.343 &0.034 &-0.001\\  
\object{NGC6121}  &-1.168 &23 & 0.179 &0.083 &53 & 0.008 &0.070 &-1.310 &0.049 &-1.302 &0.045 & 0.008\\  
\object{NGC6171}  &-1.033 &10 &-0.013 &0.058 &16 & 0.010 &0.087 &-1.072 &0.053 &-1.054 &0.072 & 0.018\\  
\object{NGC6218}  &-1.330 &18 & 0.000 &0.047 &43 &-0.024 &0.053 &-1.302 &0.028 &-1.302 &0.041 & 0.000\\  
\object{NGC6254}  &-1.575 &31 &-0.011 &0.058 &48 &-0.009 &0.064 &-1.550 &0.072 &-1.552 &0.057 &-0.002\\  
\object{NGC6388}  &-0.441 &12 & 0.007 &0.097 & 6 &-0.035 &0.062 &-0.427 &0.115 &-0.389 &0.061 & 0.038\\  
\object{NGC6397}  &-1.988 &   &	  &	 &   &       &      &	    &	   &-2.006 &0.067 &	 \\
\object{NGC6441}  &-0.430 & 3 &-0.062 &0.084 & 3 & 0.047 &0.043 &-0.284 &	   &-0.311 &	  &-0.027\\
\object{NGC6752}  &-1.555 &21 &-0.037 &0.053 &57 & 0.027 &0.055 &-1.549 &0.055 &-1.563 &0.053 &-0.014\\  
\object{NGC6809}  &-1.934 &16 &-0.035 &0.062 &55 & 0.017 &0.053 &-1.946 &0.042 &-1.955 &0.057 &-0.009\\  
\object{NGC6838}  &-0.832 &11 &-0.001 &0.042 &19 &-0.005 &0.041 &-0.803 &0.044 &-0.808 &0.038 &-0.005\\  
\object{NGC7078}  &-2.320 &12 &-0.029 &0.032 &18 & 0.003 &0.069 &-2.312 &0.066 &-2.320 &0.050 &-0.008\\  
\object{NGC7099}  &-2.344 &11 &-0.039 &0.048 &13 & 0.023 &0.059 &-2.313 &0.046 &-2.340 &0.053 &-0.027\\  
\hline
\end{tabular}
\label{t:offset}
\end{table*}

\begin{figure}
\centering
\includegraphics[scale=0.43]{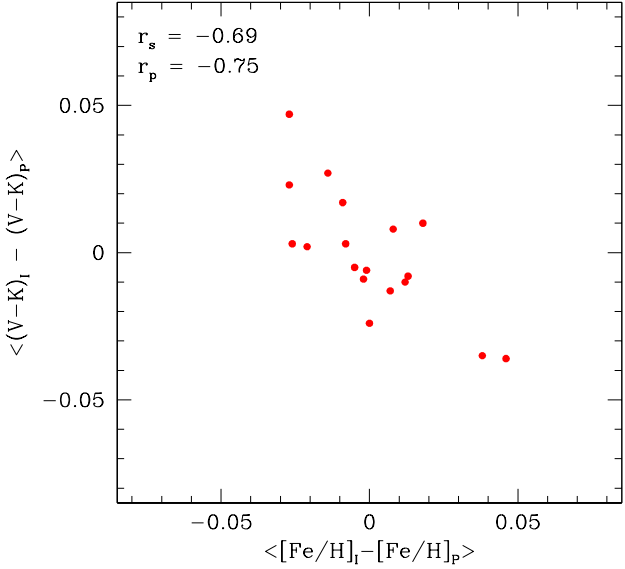}  
\caption{Mean differences in \vk \ colours between the I and P components
in the 19 GCs as a function of the metallicity difference (I-P). 
The two are well correlated, as indicated by the Pearson and Spearman
coefficients.}
\label{f:dfedvk}
\end{figure}

\begin{figure}
\centering
\includegraphics[bb=30 160 380 680 ,clip,scale=0.6]{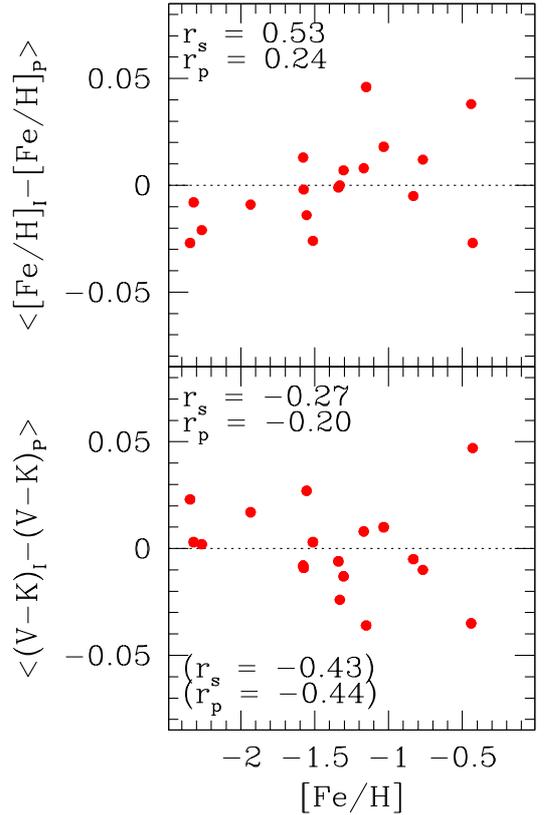}   
\caption{Individual differences I$-$P of metallicity (upper panel) and
colours (lower panel) versus [Fe/H], with Spearman and Pearson coefficients indicated
(in the lower panel, values in parenthesis are  obtained
 excluding NGC~6388 and NGC~6441).}
\label{f:fedfe}
\end{figure}

\subsection{$\Delta$Y from difference in colours}\label{deltacol}

For all stars in our 19 GCs we have Johnson $V$ magnitudes. The photometry
was obtained as described in Papers I to VIII of our series:\footnote{
In particular: in Paper I for NGC~2808, Paper II for NGC~6752, Papers  III, V for
NGC~6441, Paper IV for NGC~6218, Paper VI for NGC~6838, Paper VII for NGC~104,
NGC~288, NGC~1904, NGC~3201, NGC~4590, NGC~5904, NGC~6121, NGC~6171, NGC~6254,
NGC~6397, NGC~6809, NGC~6838, NGC~7078, and NGC~7099} 
\cite{carretta06,carretta07a,carretta07b,carretta07c,carretta09a,carretta09b} 
and \cite{gratton06,gratton07}, where the interested
reader can find  references to the original papers or description for the
unpublished data. 
We have retrieved $K$ magnitudes from the 2MASS\footnote{The Two Micron All Sky Survey
is a joint project of the University of Massachusetts and the Infrared
Processing and Analysis Center/California Institute of Technology, funded by
the National Aeronautics and Space Administration and the National Science
Foundation. } Point Source Catalog \citep{2MASS} for almost all stars. 
Distance moduli and colour excesses $E(B-V)$ were
taken from the on-line version of the GC catalog by \cite{harris}.

In our series of papers devoted to the study of Na and O abundances for 
RGB stars, we derived the temperatures of the stars using the \cite{alonso99}
relations and the photometry. In particular, to decrease the
internal error, we derived the temperatures not directly from the $(V-K)$
colours, but from relations of temperature with the $V$ or $K$ magnitude (see Paper II to VIII; 
the exception is NGC~2808, the first cluster analysed, for which we used
the classical relation between colour and temperature). 
In the present paper we used the  unreddened \vk \ colours.
The second step provided the necessary elimination of offsets between the
photometries: in each cluster we selected stars belonging to the P component, 
computed the average offset in colour between
\vk \ observed and derived from the colour-temperature relations and used it
to define the zero point for the P stars (so that they have zero colour
by definition). 

Only after this normalisation we
computed the offsets in colours between the three populations. We have
319, 587, and 39 stars in the P, I, E components, respectively. We only considered 
stars fainter than $M_K=-3.5$ because for brighter
stars the different sequences for different He are too close and
become indistinguishable\footnote{This has also the effect of minimising the possible
contamination of AGB stars, since at that magnitude level the separation between RGB and
AGB is clear. As an example, in the case of NGC~6752 , $M_K=-3.5$ means V$\sim$12.4
\citep[see][]{carretta07a}. From Fig.~1 of that paper, the separation in colour ($B-V$) between
the two sequences is then
about 0.1 mag, i.e., much more than the photometric error.}.

Results for the 19 GCs are given in Table~\ref{t:offset}, where we
indicate the cluster, the metallicity (Col. 2, taken from Paper VIII), 
the number of stars in the P component (Col. 3) and their
average colour $(V-K)_P$ and r.m.s. (Cols. 4, 5), the number of stars in the 
I component (Col. 6) and their average offset with respect to the P stars, with 
r.m.s. (Cols. 7, 8); Cols 9 to 12 show the average metallicities of the P and
I components, with their r.m.s., and Col. 13 shows the difference in [Fe/H]
(see next Section).
We did not evaluate the corresponding individual cluster values 
for the E component since the number of E stars is very small and they
would not be significant.
Figure~\ref{f:dfedvk} shows that differences in colours and in metallicity are
well anticorrelated as indicated by the Pearson and Spearman rank coefficients; this
is significant at better than the 99.9\% level. 

To see whether there is a difference in colour among the different populations
we summed up the P, I, and E components in all clusters and obtained the following weighted means\footnote{
The values would become $-0.003\pm0.004$ and $-0.042\pm0.008$ had we eliminated
NGC~3201, a cluster with differential reddening (which we corrected for anyway, 
using the maps by \citealt{von}). Its potential 
influence on the global properties of our sample
is larger than the one by e.g., NGC~6388 or NGC~6441 because we have a much
larger number of stars. Even so, the results are indistinguishable.}:
$$ (V-K)_I - (V-K)_P = -0.003 \pm 0.004 ~~(r.m.s.=0.060) $$
$$ (V-K)_E - (V-K)_P = -0.037 \pm 0.010 ~~(r.m.s.=0.056) $$
We see that $\Delta(V-K)$ is of the same order of the error, hence not 
significant, for the P and I stars, while it is about 4 times the error for
the P and E stars. We conclude that the extreme populations is significantly
bluer than the primordial one. This $\Delta(V-K)$  corresponds to a
$\Delta$Y of about 0.05 or 0.08, according to the models cited, for case $a$)
and $b$), respectively. The differences in colour and in metallicity (see Sec.
2.3) are summarised in Table~\ref{t:delta}, while the corresponding $\Delta$Y
values are shown in Table~\ref{t:deltaY}, for simplicity.

A complication may arise from the presence of CO bands  in the $K$ filter. They
are stronger in P stars than in O-depleted I and E ones, and may depress the
flux in the $K$ filter by about 1-2\%, making the P stars to appear bluer, hence
decreasing the difference in \vk \ among populations. From simulations based on
the CO index by \cite{cpf78}  we have estimated that the
\vk \ of P stars should be about 0.013 mag redder than they appear. That means
that the corrected $\Delta(V-K)$ between E and P is about -0.05; 
this corresponds to $\Delta$Y of about 0.06 or 0.10, for the two mixtures.

This is a global result, averaged over the 19 GCs; it is interesting to consider
NGC~2808 separately, the
only cluster in our sample for which large differences in He have been
deduced  on the basis of the multiple MSs \citep{piotto07}. 
We find for NGC~2808:
$$(V-K)_I - (V-K)_P = -0.036 \pm 0.015,$$
$$(V-K)_E - (V-K)_P = -0.044 \pm 0.017$$ 
For this cluster both the I-P difference and the E-P one
are significant at about the  2.5$\sigma$ level.
These $\Delta(V-K)$ imply a $\Delta$Y of about 0.05 (case $a$) or 0.08 (case $b$),
and 0.08 (0.09)  between the P and I or E populations, respectively.
Taking again into account the effect of CO bands, these differences
become $\Delta(V-K)_{I-P}=-0.049$ and $\Delta(V-K)_{E-P}=-0.057$ which
translate into  $\Delta{\rm Y}_{I-P}=0.06 ~(0.10)$ and 
$\Delta{\rm Y}_{E-P}=0.07 ~(0.11)$,
respectively.\footnote{The small difference in $\Delta{\rm Y}_{E-P}$ between
the average of 19 GCs and NGC~2808 alone is due to the fact that the latter has
a strong contribution to the global E population.}

\begin{table}
\centering
\caption{Summary of differences in colour and metallicity between populations.}
\setlength{\tabcolsep}{1.0mm}
\begin{tabular}{l cccc}
\hline\hline
Sample & $\Delta(V-K)_{IP}$  & $\Delta(V-K)_{EP}$
              & $\Delta{\rm [Fe/H]_{IP}}$  & $\Delta{\rm [Fe/H]_{EP}}$ \\
\hline
 all           &  -0.003 &-0.037  & 0.000 & 0.027 \\
 all(+corr)$^1$      &    -    &-0.050  &   -   &    -  \\
NGC~2808       &  -0.036 &-0.044  & 0.046 & 0.084 \\
NGC~2808(+corr)$^1$  &  -0.049 &-0.057  &   -   &    -  \\
\hline
\end{tabular}
\tablefoot{$^1$ With correction for the CO bands.}
\label{t:delta}
\end{table}

\subsection{$\Delta$Y from difference in d$\theta$}\label{deltatemp}

In alternative to the use of temperatures derived from colours, we may also use
information derived from excitation  to find  (small) temperature differences
between the three populations.\footnote{
The definition  
is: $\theta_{exc} = 5040/T_{exc}$. Since $T_{exc}\simeq 0.86\times T_{eff}$, we
have  $\theta_{exc} = 5860/T_{eff}$. Differentiating the equation
we have the relation between d$\theta$ and d$T_{eff}$.
}
This is possible because we did not derive $T_{eff}$'s from the excitation
equilibrium (i.e. the spectroscopic route) but from the photometry.
From the excitation equilibrium we derived the d$\theta$ values for all the
programme stars; recalling that d$\theta>0$ means that the star is warmer than
expected from its luminosity, we found the following weighted average values:
$ d\theta_E-d\theta_P = +0.0092 \pm 0.0040 $
using the 39 E stars and the 207 P stars with more than 25 lines measured.
The result improves to
$ d\theta_E-d\theta_P = +0.0094 \pm 0.0037 $
if we relax the request on the number of lines and use the 61 E stars and the 
315 P stars with more than 15 lines measured.
This difference is significant at 2.5$\sigma$ and corresponds,
at 4500 K, the mean temperature of the
stars considered, to a difference
in temperature of  $32 \pm 14 $ K, in the sense of  E stars being warmer than P ones. 
This implies a $\Delta$Y of about 0.06 between E and P stars (an average between results
for the cases $a)$ and $b)$ discussed above).
Since in this case we have averaged stars of all magnitudes and the sequences
tend to converge at brighter luminosity, the average difference is smaller than
the one implied by colours.

Once again, the difference between the P and I populations is not significant;
the corresponding differences are $ d\theta_I-d\theta_P = -0.0037 \pm 0.0017 $
and $ d\theta_I-d\theta_P = -0.0018 \pm 0.0018 $ considering stars with at least
25 or 15 lines measured, respectively.

\begin{table*}
\centering
\caption{Summary of differences in Y between first and second-generation stars.  }
\setlength{\tabcolsep}{1.4mm}
\begin{tabular}{l cccc c cccc}
\hline\hline
Sample & $\Delta {\rm Y_{PI}}(V-K)$  & $\Delta {\rm Y_{PE}}(V-K)$
       & $\Delta {\rm Y_{PI}}[Fe/H]$  & $\Delta {\rm Y_{PE}}[Fe/H]$ 
       &
       & $\Delta {\rm Y_{PI}}(V-K)$  & $\Delta {\rm Y_{PE}}(V-K)$
       & $\Delta{\rm Y_{PI}}[Fe/H]$  & $\Delta {\rm Y_{PE}}[Fe/H]$ \\
\hline
\multicolumn{1}{c}{} &\multicolumn{4}{c}{Case $a$} & &\multicolumn{4}{c}{Case $b$} \\ 
 \cline{2-5} \cline{7-10} 
 all	      &   ...   & 0.049 &  ...	& 0.046 & &  ...   & 0.077  &  ... & 0.11 \\ 
 all(+corr.)$^1$     &   ...   & 0.064 &  ...	&   ...  & &  ...   & 0.100  & ...   &  ...  \\ 
NGC~2808      & 0.047 & 0.057 & 0.109   & 0.146 & &  0.075 & 0.089  & 0.14  & 0.19 \\ 
NGC~2808(+corr.)$^1$ & 0.063 & 0.072 & ...	&  ...   & &  0.098 & 0.113  & ...&  ... \\ 
\hline
\end{tabular}
\tablefoot{$^1$ With correction for the CO bands.}
\label{t:deltaY}
\end{table*}

\subsection{$\Delta$Y from difference in [Fe/H]}\label{deltamet}

Grossly speaking, given the definition X+Y+Z=1, for a fixed global metallicity
Z, an increase in He (Y) has to be associated to a decrease of the H abundance,
so that  [Fe/H] increases. Of course, the issue is more complex, since a change
in the H to He ratio has effects on the star structure, which are taken into
account by the evolutionary models used in the present paper.

We noticed this feature in NGC~2808 \citep{carretta06} when,  dividing stars in
a way different from the one adopted here, we found slightly different (and
increasing) values of [Fe/H] for ``O-normal",  intermediate, and very O-poor
stars (i.e., with canonical, intermediate, and enhanced He in the usual
interpretation of Na-O anti-correlation). We explore here the whole sample of
our programme clusters.

In this case we have to proceed in a slightly different way. As mentioned in
Sect. 2.1, in our series of papers (Paper II to Paper VII, but notably not in
Paper I on NGC~2808) we derived the temperatures used to obtain [Fe/H] and other
abundance ratios from a calibration between magnitude and colour-based
temperature. This reduced the effect of errors (since magnitudes are more
reliably measured than colours) and allowed to have smaller internal errors, but
also effectively collapsed all possible differences in metallicity due to
different He, that would have shown as different colours, hence temperatures.
Before comparing [Fe/H] values of the different populations we have then applied
a correction to transform the metallicities to the values they would have had,
had we used directly the colour-based $T_{eff}$'s.

When we apply the same procedure
adopted for colours, i.e. normalising to the P
population of each cluster (for a total of 320 stars), we obtain the
following weighted means:
$${\rm  [Fe/H]_I - [Fe/H]_P =  0.000\pm0.003   ~~(r.m.s=0.051)} $$
$${\rm [Fe/H]_E - [Fe/H]_P =  0.027\pm0.010   ~~(r.m.s=0.059)} $$
with 587 stars in the I and 38 stars in the E component, respectively.
If we eliminate NGC~3201  because of its differential reddening,
things do not change much: these two
numbers become $0.002\pm0.004$ (288 stars, r.m.s=0.055) and $0.032\pm0.010$ 
(32 stars, r.m.s.=0.059), respectively.
Only the difference between E and P populations is significant and, for the
two cases, it
implies a $\Delta{\rm Y}\simeq0.05$  (or 0.06, without NGC~3201) if the whole
difference is due to the denominator in [Fe/H], i.e. if the distribution of
heavy elements is the same in P and E stars (case $a$). The situation is more
complicated in case $b$, where also Z changes; we have run models with 
different assumptions for the change in Z (from zero to 0.002); a
$\Delta$[Fe/H]=0.03 implies for an average difference in Z of 0.0013, a
$\Delta{\rm Y}\simeq 0.11$. 

Also in this case we may separate NGC~2808 and we obtain 
$${\rm   [Fe/H]_I - [Fe/H]_P =  0.046\pm0.020   ~~(r.m.s=0.066}) $$
$${\rm   [Fe/H]_E - [Fe/H]_P =  0.084\pm0.018   ~~(r.m.s=0.052}) $$
Both values are significant and imply $\Delta{\rm Y}_{I-P}=0.11$ and
$\Delta{\rm Y}_{E-P}=0.15$ (case $a$).
For case $b$, these values become, again assuming the same difference in Z as
above,
$\Delta{\rm Y}_{I-P}=0.14$ and
$\Delta{\rm Y}_{E-P}=0.19$

\begin{figure}
\centering
\includegraphics[scale=0.4]{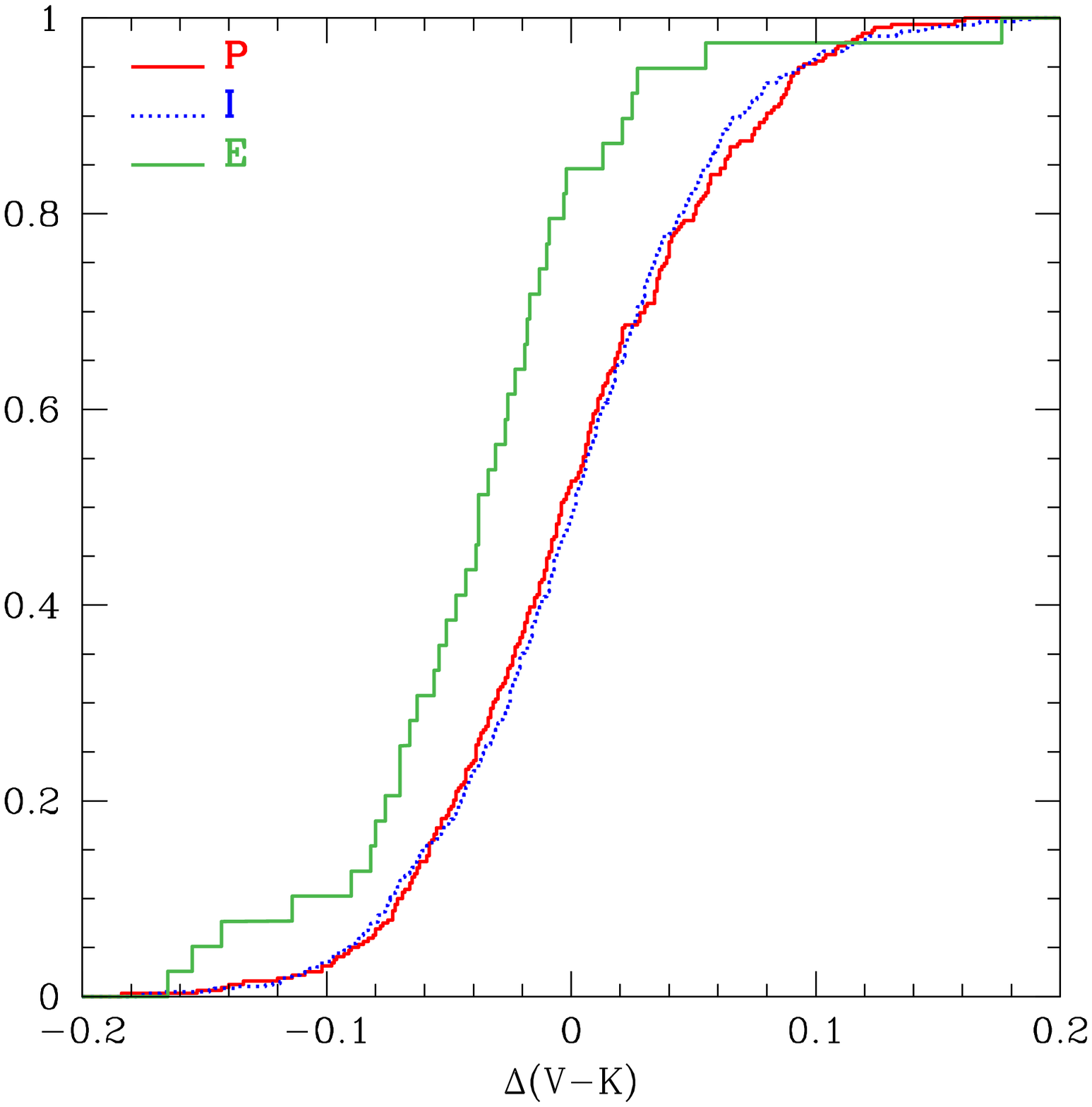}
\includegraphics[scale=0.4]{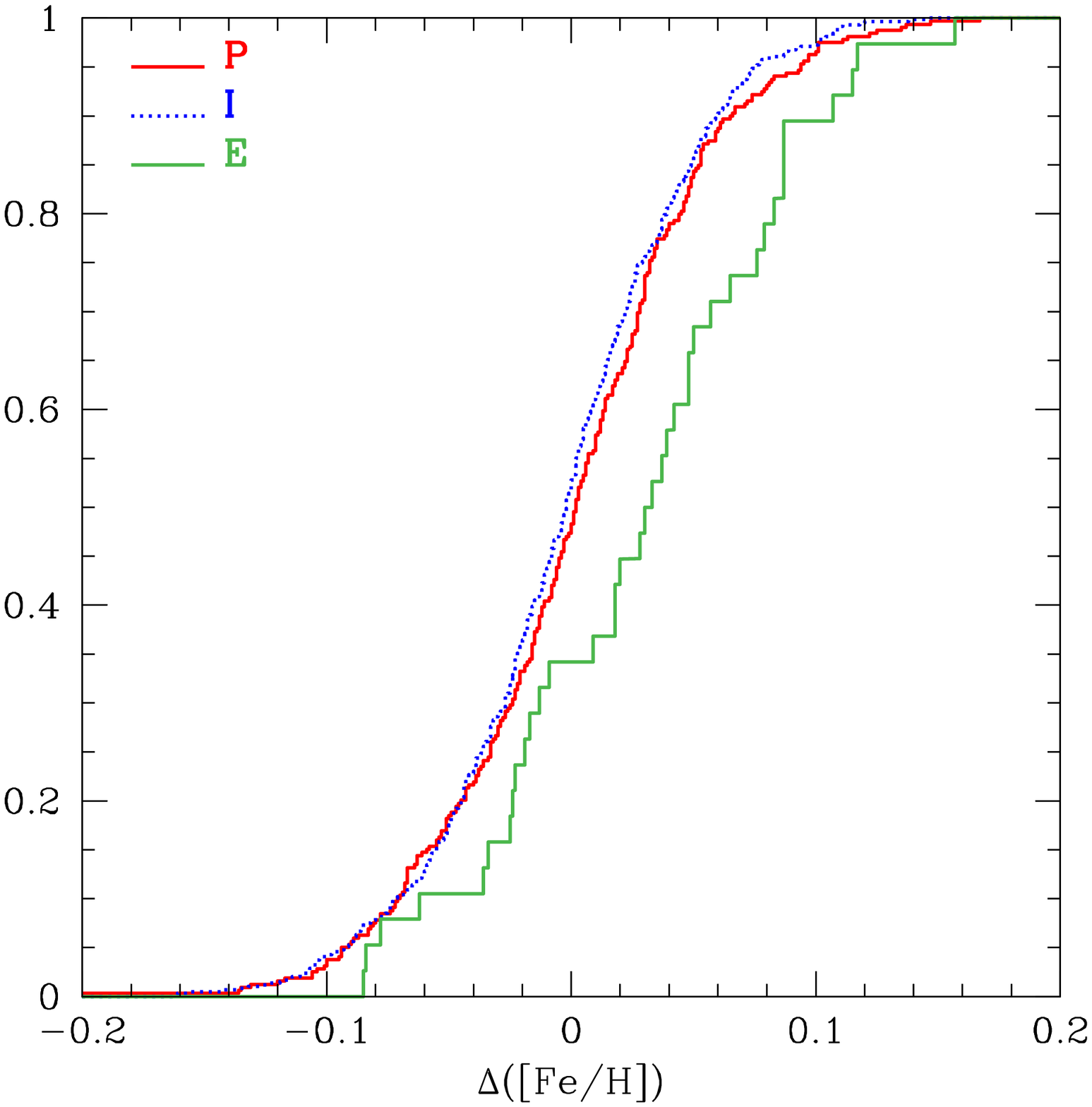}
\caption{Cumulative distributions for difference in colour (upper panel) and
metallicity (lower panel) for individual stars in the three populations.}
\label{f:cumu}
\end{figure} 

\subsection{Differences among the three populations}\label{intermediate}

Even if this  is part of our global study of the
anticorrelations in GCs, we have to remember that $\Delta$Y is not a direct
synonym of anticorrelation and that the same extension of the anticorrelations
(as measured, e.g. from the Interquartile range IQR[O/Na], see 
\citealt{carretta06,carretta07d}), may be reached for different values of $\Delta$Y.
We have an indication of that from Fig.~\ref{f:fedfe}: in the
lower panel we plot the difference in metallicity and colour between I
and P stars as a function of cluster metallicity, while in the upper panel we do the same
for differences in metallicity. The individual values  have large  error (and this is why  
we considered
only the averages over the whole sample); however,  there is a hint that the effects of
differences in He content (as inferred from the difference in metallicity or colour) are less 
evident for  the metal-poor clusters. Only in the metal-richer clusters the I stars are
bluer and metal-richer than the P ones. This could suggest that, for the same
production of proton-capture elements (e.g., destruction of O and production of
Na), more He is produced in metal-richer clusters. In other words, in the
metal-poor GCs a smaller $\Delta$Y is required to produce an effect on the shape
of the Na-O anticorrelation, i.e., on the classification of a star in the I or
even E population. 

While the effect discussed above needs corroboration,  a solid conclusion is that,
with the exception of NGC~2808, the differences in colour and
metallicity are significant only between the extremes of the GC populations.
This can be immediately visualised using the cumulative distributions of the differences in colour
and metallicity for the individual stars (i.e., the difference between each star and the
average of the P population). Fig.~\ref{f:cumu} shows these cumulative distributions,
separated for P, I, and E stars. The P and I distributions look very similar in both panels, 
while the E stars show a distinct behaviour (the probability that they are taken  from the
same parent distribution of the
P stars are very low: 4$\times10^{-4}$ and 0.015 for colour and metallicity, respectively): 
as a whole, they
are bluer and metal-richer.

\section{He from the luminosity function of Na-poor and Na-rich
stars}\label{bump}

\begin{figure}[h!]
\centering
\includegraphics[bb=70 180 570 680, clip,scale=0.35]{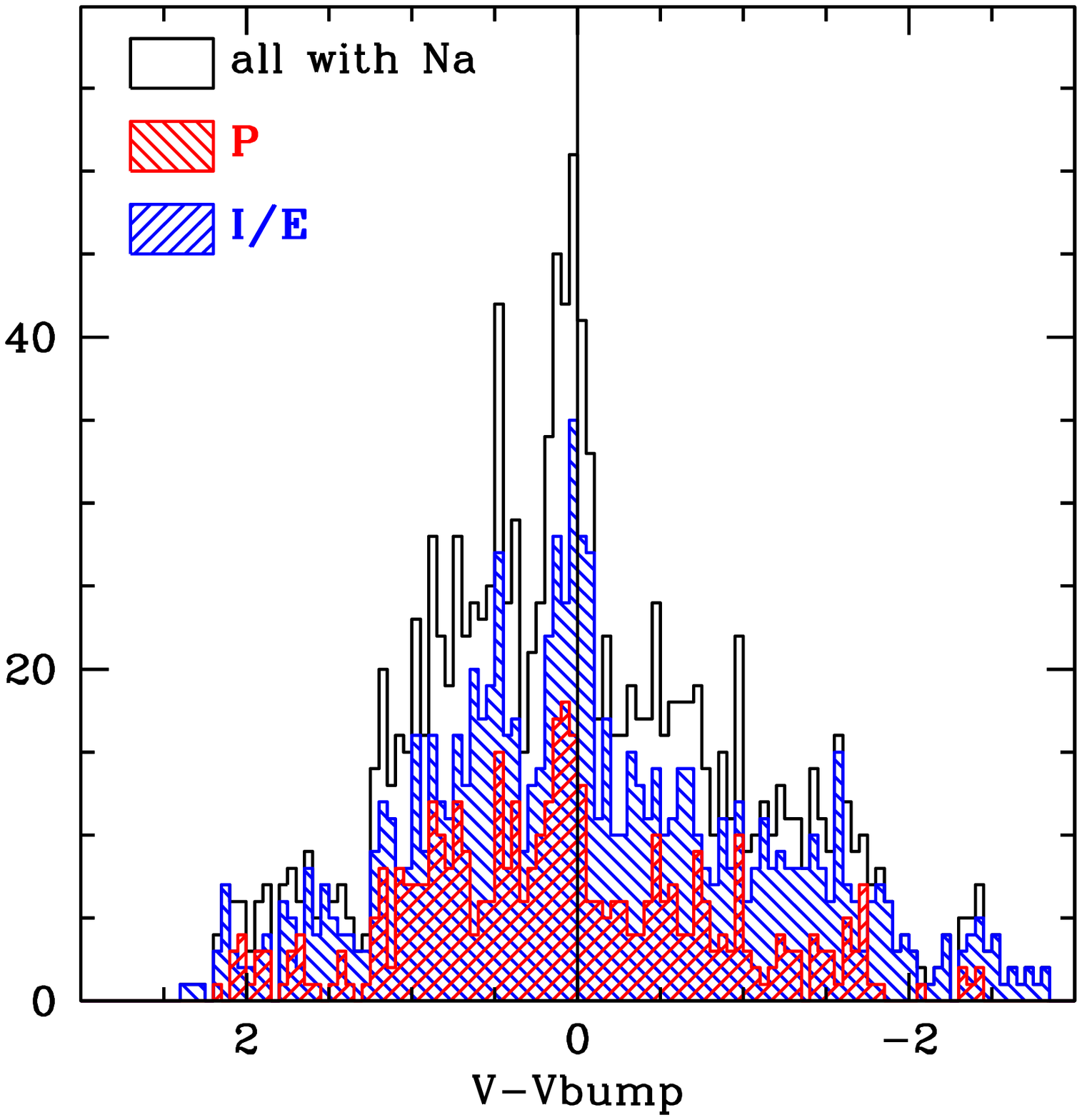}
\includegraphics[bb=80 180 380 720, clip, scale=0.6]{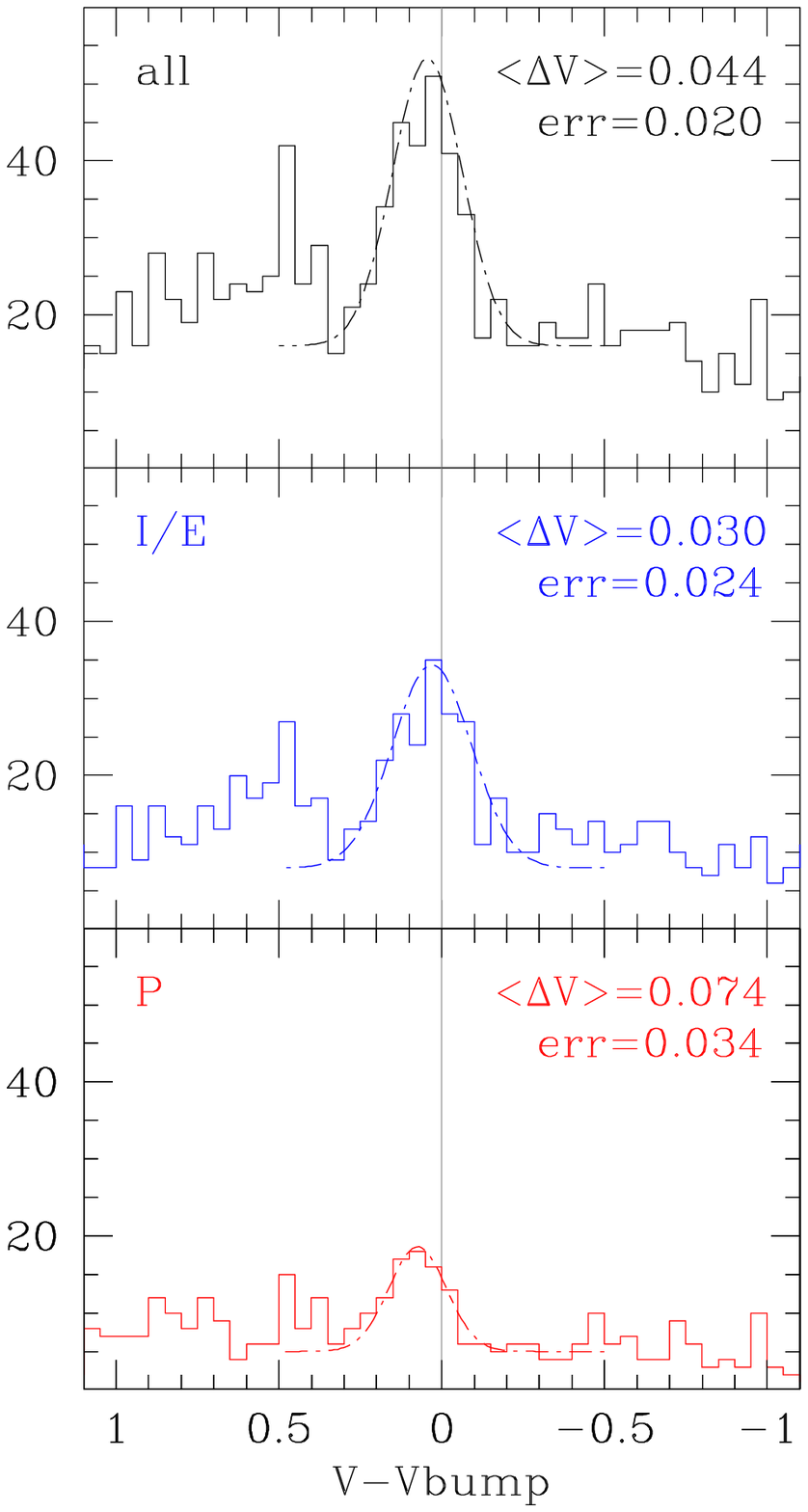}
\caption{Upper panel: Histogram of the difference in magnitude between
individual $V$ and  $V_{bump}$ (for each cluster) for all stars with Na measured
(open histogram) and for first-generation (red histogram) or second-generation
(blue histogram). Lower panels: Zoom near the RGB bump for all, second
generation (I and E), and first-generation  stars, with Gaussians fitting each
bump. In each panel  the peak of the bump and the associated error are
indicated. }
\label{f:istobump}
\end{figure}

\begin{figure}[htb]
\centering
\includegraphics[scale=0.35]{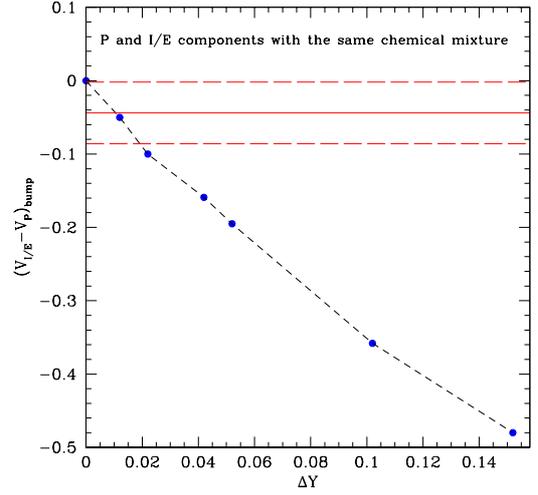}  
\caption{The V-band RGB bump brightness difference between the P sub-population and
the I/E one as a function of the He abundance difference between the two
sub-populations. In this case, for both sub-populations it has been assumed the
same {\sl normal}, $\alpha-$enhanced heavy elements distribution (case $a$ in
Sect.~\ref{sec2}). The solid line represents the measured empirical RGB bump
luminosity difference between the P and I/E sub-populations, while the
long-dashed lines shown the same value at $\pm1\sigma$ level.}
\label{bumpteo1}
\end{figure}

\begin{figure}[htb]
\centering
\includegraphics[scale=0.35]{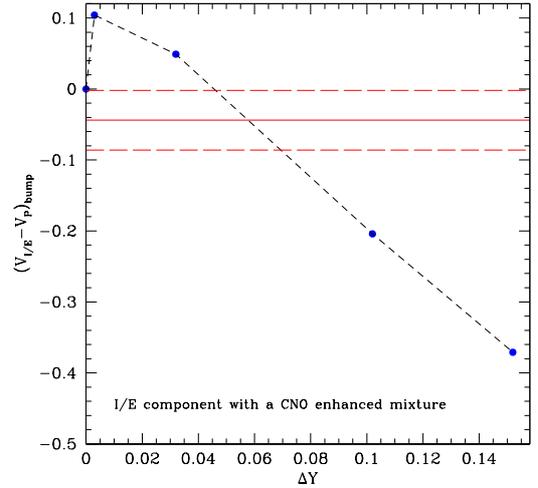}  
\caption{As in Fig.\ref{bumpteo1}, but taking into account the fact that the I/E
sub-population is characterised by a peculiar chemical pattern with a CNONa
anticorrelation  (case $b$).}
\label{bumpteo2}
\end{figure}

With the intent of finding an additional evidence supporting our scenario for
the correlation between Na-rich stars and their initial helium content, we
tested also the possibility to detect any brightness difference between the RGB
bumps corresponding to the two distinct stellar populations. In fact, canonical
stellar evolution predicts that the RGB bump brightness increases when the
initial He content increases \citep[see, for instance][]{salcas}. As a
consequence, one should expect to observe this occurrence when comparing the
primordial stellar component with the intermediate/extreme one
\citep[e.g.,][]{salaris06}.

However, the possibility to detect this observational signature in the
luminosity function (LF) of a given GC is generally strongly hampered by the
inadequate statistics (the number of RGB stars in the various LF bins is not
large enough). We made a first explorative attempt  (Paper IV) combining the
samples in NGC~6218 and NGC~6752 to improve the statistics. Here, in order to
better overcome this problem, we decided to combine the data for the LFs
corresponding to the various GCs in our database. We took the values for
$V_{bump}$ from literature (see Table~\ref{t:bump}) but could apply the test
only to 14 of the 19 GCs. We did not reach the level of the bump in some of our
target clusters, either  because they were very far, like NGC~6388, NGC~6441  or
because they were so rich, like NGC~104, NGC~2808, that we had enough target in
the upper RGB and did not need to reach down along the RGB (our original goal
was  to obtain good spectra to measure Na and O abundances for about 100 stars
per cluster,  so we selected preferentially bright targets). Finally, we did not
find a $V_{bump}$ value for NGC~7099; since it has the same metallicity of
NGC~7078, we could have adopted the value for the latter, taking into account
the distance moduli, but we would have introduced an additional uncertainty,
given also the slight difference in age  (\citealt{carretta09c}; note that these
two GCs have the same age in \citealt{marin09}).

We have in total 1368 stars in the 14 GCs with a measured Na abundance. To
separate P stars from I and E ones we used [Na/Fe]$_{min}+0.3$
\citep[see][]{carretta09a}; we have 438 P stars and 930 I, E ones (we sum the
second-generation populations, since the E stars are a minority). We computed
for each cluster the residuals between the individual $V$ values and $V_{bump}$,
to eliminate the dependence on metallicity and cluster age.  We produced
histograms both for the whole sample, and for the P and I/E stars separately.
The upper panel of Fig.~\ref{f:istobump} shows the resulting luminosity
functions. Since we have here only the stars for which we obtained a spectrum
and not the entire RGB, the part below the bump is strongly incomplete and we do
not see the usual increase going towards fainter stars found in photometric
works on the RGB  luminosity function. However, the bump is visible both in the
whole sample and in the two sub-samples which show slightly different peak
values. The small offset from zero in the histograms is due to the the different
photometries adopted in the definition of $V_{bump}$ and in our programme (see
below).

To better appreciate this, we present in the lower panels of
Fig.~\ref{f:istobump} a zoom in the bump region, separating the three
components. Fitting a Gaussian to the bumps we derive a difference in magnitude
between P and I, E stars of 0.044 mag, in the direction expected from theory,
with I, E stars presenting a brighter RGB bump than P stars. Unfortunately, the
significance is low, because of the errors on the position of the peaks. A first
source of error is the statistics: the bump peak (i.e., above the background) is
produced by only about 150 stars, two thirds of which in the I, E population and
one third in  the P one. To halve the errors in the position of the two separate
peaks we should increase the number of stars by at least 3 times and this is not
an immediate possibility because of the large amount of telescope time and
analysis implied. Other sources of errors come from the photometry: i)
$V_{bump}$  is measured at best with an uncertainty of 0.05 mag, ii) the
magnitudes we use have a small attached error, and iii) we did use a different
photometry from the one on which $V_{bump}$ was determined (although the
possible offsets  should be amply accommodated within the uncertainty on
$V_{bump}$). However, even if we tried to re-determine the bump positions in our
photometric data, the situation would not become significantly better. We note
in fact that the bump has an intrinsic width of the order of 0.2 mag
\citep{salaris02} and that we have already reached that level: the FWHM of the
Gaussian used to fit the P population (the one with a single Y value) is 0.21
mag (while the FWHM of the second-generation stars, with varying Y content, is
larger, about 0.28 mag).

In order to estimate the average difference in the initial He content between
the P stellar population and the I/E one, from the measured RGB bump brightness
difference, we used the extended set of low-mass stellar models for various
assumptions about the initial He content described in Sect.~\ref{sec2}.  

As a first step, it has been assumed that the P and I/E stellar populations have
exactly the same heavy elements distribution (case $a$); then for a fixed age of
$\sim12.5$ Gyr and iron content [Fe/H]=$-1.31$, we have computed the synthetic
LF\footnote{In order to take into account the observational errors that can
affect the photometric data for each GC in our sample, when computing the
synthetic LFs we have considered a photometric error of $\sim0.02$ mag.} 
for the P population by using a canonical He abundance (Y=0.248), whereas for
the I/E population various synthetic LFs have been computed by using various
He-enhanced abundances in the range from 0.26 to 0.40. By using these
theoretical LFs, we have estimated the brightness difference between the RGB
bump of the P population and that of the I/E one as a function of the difference
in the initial He contents of the two populations. The results of this analysis
are shown in Fig.~\ref{bumpteo1}. The comparison between the model predictions
and the estimated empirical difference suggests a difference in the mean He
content between the P and the I/E sub-populations of about $0.01\pm0.01$.

Case $b$ should represent a more significant comparison between theory and
observations; the adopted heavy elements distribution corresponds to a mixture
where the sum (C$+$N$+$O) is enhanced by about a factor of 2 with respect to the
{\sl reference}, $\alpha-$enhanced mixture. This value is consistent - although
it represents an upper limit - with the results of the spectroscopical analysis
performed by \cite{carretta05} for the extreme values of the chemical
anti-correlations observed in GCs. The RGB bump brightness difference between
the P sub-population and the I/E one, when for the latter a CNONa peculiar
mixture is assumed,  is shown in Fig.~\ref{bumpteo2} as a function of the He
abundance difference. This figure deserves some comments: the RGB bump
brightness of the I/E sub-population is fainter than that of the P
sub-population for an He content difference lower than about 0.045. This
occurrence is due to the fact that when comparing $\alpha$-enhanced stellar
models with those accounting for a CNONa peculiar pattern, we are considering
the same iron content [Fe/H], and this means that the global metallicity of the
stellar models with the peculiar heavy elements distribution is larger than that
of the {\sl reference} models. Therefore, as discussed by 
\cite{pietrinferni09}  the RGB bump - at fixed age and [Fe/H] value - is
expected to be fainter in these sub-populations characterised by a peculiar
chemical pattern  \citep[see also][]{salaris06}. However, for increasing
difference in the He content, the effect of the He abundance on the RGB bump
brightness overcomes that associated to the larger metallicity, and the bump of
the I/E sub-population becomes brighter with respect the P sub-population.
Theoretical models shown in Fig.~\ref{bumpteo2} suggest an average difference
between the P and the I/E sub-population of $\sim0.06\pm0.015$. This is in
better agreement with the estimates obtained in previous sections from
independent indicators, even if lower. Consider however that NGC~2808 is not in
the sample and that this cluster brings a strong signature of Y enhancement.

As stated above, all the stellar models used in the present analysis have
been computed by assuming a constant [Fe/H] value regardless of the
adopted initial He content.   It is in fact  expected that both candidate polluters
(AGB and FRMS)  do not alter the
initial iron content. However,  it could be worthwhile to check at what extent - if any -
the present results are affected by this assumption. For this purpose, we have 
computed suitable stellar models for selected initial He contents, keeping the iron 
content fixed. We found that the magnitude difference between the RGB bump
of the I/E sub-population and that of the P one
is increased by a negligible 0.01 mag for a He difference between the two
sub-populations equal to $\Delta{Y}=0.05$ and of about 0.09 mag for
$\Delta{Y}=0.10$. From  Figs. \ref{bumpteo1} and \ref{bumpteo2}, it is clear that
this change does not affect  our previous estimates of the He
content difference.

Finally, we wish to draw attention on a possible test; clusters with larger
$\Delta$Y should also display broader RGB bumps. From Figs.~\ref{bumpteo1},
\ref{bumpteo2} we estimate that a $\Delta$Y=0.1  should correspond to a visible
broadening of the bump (since first and second-generation stars should display
differences of about 0.20-0.35 mag between their respective bumps).  Precise
photometry of all their RGB stars should be used. An ideal couple for this
comparison are the two GCs NGC~2808 and NGC~6121 who have similar metallicity
but very different extension of the Na-O anticorrelation (Carretta et al, paper
VII) and very different implied  $\Delta$Y \citep{gratton10}. 

\begin{table}
\centering
\caption{Adopted values of $V_{bump}$.}
\begin{tabular}{cccc}
\hline\hline
GC & $V_{bump}$  & $\pm$ & Reference \\
\hline
NGC~0288 &15.45 &0.05 & 1 \\
NGC~1904 &16.00 &0.04 & 2 \\
NGC~3201 &14.55 &0.05 & 1 \\
NGC~4590 &15.15 &0.05 & 1 \\
NGC~5904 &15.00 &0.05 & 1 \\
NGC~6121 &13.40 &0.10 & 1 \\
NGC~6171 &15.85 &0.05 & 1 \\
NGC~6218 &14.60 &0.07 & 1 \\
NGC~6254 &14.65 &0.05 & 1 \\
NGC~6397 &12.60 &0.10 & 3\\
NGC~6752 &13.65 &0.05 & 1 \\
NGC~6809 &14.15 &0.05 & 1 \\
NGC~6838 &14.80 &0.15 & 1 \\
NGC~7078 &15.25 &0.05 & 1 \\
\hline
\end{tabular}
\tablebib{
(1) \cite{ferraro99}; (2) \cite{zoccali99}; (3) \cite{as99}
}
\label{t:bump}
\end{table}

\section{Discussion and summary}\label{ultima}

Recently, spreads and splits in GC RGBs other than the ones  (in $V-K$,
metallicity and temperature) considered here have been presented. They are most
probably due to the same phenomenon we are seeing here, since the He-enhanced
stars should also show other chemical signatures. For instance, the correlated
N-enhancement should be responsible for the effects seen in the colour-magnitude
diagrams involving the Johnson $U$ or Str\" omgren $u$ filters
\citep{marino08,yong08}. On the other hand, the effect seen in  $hk$ photometry,
attributed to a  spread in calcium by \cite{lee09}, and demonstrated not to be
so by \cite{carretta10}, still awaits a definite explanation. 

We have used information on RGB stars in GCs to infer the plausible He
differences implied by the existence of two generations of stars. This is a new
approach, since the different methods used in the past to deduce the He content
of GCs generally involve the HB,  the evolutionary phase more sensitive to even
small He variations (the two exceptions being $\omega$~Cen  and NGC~2808 with
their multiple MSs). 

Apart from the information coming from photometric data, He can be deduced from
spectra of HB stars, where He lines are visible for temperatures hotter than
about 8500 K. \cite{villanova09}  studied seven HB stars in NGC~6752, using UVES
spectra at high-resolution and very high S/N. For the four stars where they were
able to measure the He line, they obtained a value similar to the cosmological
one (Y=0.245). However, these four are all O-rich, Na-poor, while  Villanova et
al.  could not measure He for the only Na-rich, O-poor star, the one that
should  also be He-enhanced, so the case is not yet settled. However, the
photospheric abundances of HB stars hotter than about 11500 K are altered  by
atomic diffusion processes and cannot be used directly to measure the original
values.  For instance, Behr (2003) found a depletion in moderately hot HB stars
in several GCs. Moehler et al., in a series of papers 
\citep[e.g.,][]{moehler07} concentrated instead on the hottest part of the HB
(and blue-hook stars in the two GCs -$\omega$ Cen and NGC~2808- where they are
present) and found He-enhancement. This can be attributed either to  pollution
from a previous generation  or to  He flash induced mixing occurring in hot
He-flashers, i.e. those stars experiencing the He flash not at the RGB tip but
along the White Dwarf cooling sequence \citep[see][]{cc93}.  A further
discussion of this problem in $\omega$ Cen can be found in \cite{cassisi09}.

We have studied in a companion paper \citep{gratton10} the effect of (even a
small) difference in He content on the morphology of HBs, and found that He is
most probably the third parameter governing the HB, after metallicity and age.
We refer to that paper for a long and detailed discussion on the effects of He
variations, on methods to measure He content \citep[e.g., the
R-method:][]{iben}, on comparison between results obtained through different
approaches. In the present paper we limit our analysis to the differences in He
that can be deduced from RGB stars.

May we use the present analysis, combined with the previous work on the Na-O 
and other (anti)correlations to try discriminating between  AGB and FRMS as the
polluters of the second-generation stars? We need to consider that He is
produced essentially in MS for both  classes of polluters, while there is a
difference for the other ``peculiar" elements (O, Na, Mg, Al, Si), produced
during  Hot Bottom Burning or in MS, respectively. However, we still miss
important details of  stellar evolution (e.g., how much He is dredged up in AGB
stars, or what is the dependence on metallicity -via rotation- of FRMS), the
numbers are still small and the inferences are not  conclusive. We need to
further investigate the dependencies of variations of He and  those other
elements  on the stars and cluster properties.

In summary, with the present work we have seen that:
\begin{itemize}
\item[i)] it is possible to deduce variations of He also from RGB stars in GCs
when using large samples, treated homogeneously;
\item[ii)] these variations are generally measurable from colour (here $V-K$)
and [Fe/H] only between first-generation (P) and extreme second-generations (E)
stars;
\item[iii)] the variations implied for the average of the 19 GCs considered are
of the order of 0.05-0.10 ($\Delta$Y); however, the heavy-elements mixture
assumed for the second-generation stars has to be taken into account (and we
have presented two extreme cases);
\item[iv)] NGC~2808 is a notable exception: in this cluster, differences in 
colour and metallicity are seen also between P and the intermediate (I)
component of second-generation stars. This is perfectly in line with the fact
that NGC~2808 presents a clear evidence of three He levels also in MS. For
NGC~2808 the deduced $\Delta$Y values are higher (see Table~\ref{t:deltaY}) than
for the average of all GCs;
\item[v)] similar results can be obtained also considering the luminosity of the
RGB bump.  We could test this method only for 14 GCs; unluckily, NGC~2808 is not
among them. Within the limits of the rather poor statistics,  we found that the implied  
$\Delta$Y values are in reasonable
agreement with those found from colour and metallicity;
\item[vi)] the absolute calibration of the $\Delta$Y in GCs is still to be
defined.
\end{itemize}

\begin{acknowledgements}
This work was partially funded by  the grant
INAF 2005 ``Experimenting nucleosynthesis in clean environments"
and the PRIN MIUR 2007 ''Multiple stellar populations in Globular Clusters". 
SL is grateful to the DFG cluster of excellence ``Origin and Structure of
the Universe" for support. 
This research
has made use of the SIMBAD database, operated at CDS, Strasbourg,
France and of NASA's Astrophysical Data System. We thank the referee
for her/his comments, useful to improve the presentation.
\end{acknowledgements}

\end{document}